\documentclass[10pt]{article} 
\usepackage{spie}
\usepackage{latexsym}
\usepackage{times,mathptm} 
\usepackage{graphicx} 
\usepackage{multicol}
\usepackage{rotating} 
\usepackage[verbose]{wrapfig}
\usepackage{epsfig}
\newlength{\dinwidth}
\newlength{\dinmargin}
\newcommand{\bc}{\begin{center}}
\newcommand{\ec}{\end{center}}
\newcommand{\be}{\begin{equation}}
\newcommand{\ee}{\end{equation}}
\newcommand{\bi}{\begin{itemize}}
\newcommand{\ei}{\end{itemize}}
\newcommand{\bt}{\begin{table}}
\newcommand{\enta}{\end{table}}

\newcommand{\g}{$\gamma$}

\newcommand{\beq}{\begin{equation}}
\newcommand{\eeq}{\end{equation}}
\newcommand{\bfg}{\begin{figure}}
\newcommand{\efg}{\end{figure}}
\newcommand{\keV}{\mbox{ke\hspace{-0.1em}V}}
\newcommand{\MeV}{\mbox{Me\hspace{-0.1em}V}}

\renewcommand{\deg}{\ensuremath{^\circ}}

\graphicspath{{figures/}}
\begin{document}
\title{Preliminary Results from the 1999 Balloon Flight of the Liquid Xenon Gamma-Ray Imaging Telescope (LXeGRIT)}  
%
\author{
        E.~Aprile$^a$, U.G.~Oberlack$^a$, A.~Curioni$^a$, V.~Egorov$^a$,
	K.-L.~Giboni$^a$, \\
        S.~Ventura$^{a,b}$, T.~Doke$^c$, J.~Kikuchi$^c$, K.~Takizawa$^c$
	E.L.~Chupp$^d$, P.P.~Dunphy$^d$ \\ 	
        \skiplinehalf
        $^a$Columbia Astrophysics Laboratory, Columbia University \\
	$^b$INFN--Padova, Italy \\
	$^c$Waseda University, Japan \\
	$^d$University of New Hampshire, USA \\
}
\authorinfo{Send correspondence to: Elena Aprile, Columbia University, 
Astrophysics Laboratory, 550 West 120th Street, New York, NY 10027 \\
E-mail: age@astro.columbia.edu \hfill \\
LXeGRIT Web page: \texttt{http://www.astro.columbia.edu/$\sim$lxe/lxegrit/}
}
\pagestyle{plain}    
\maketitle

\begin{abstract}
LXeGRIT is a balloon-borne Compton telescope employing a large volume liquid
xenon time projection chamber (LXeTPC) as the central \g--ray detector. It is
designed to image \g--rays in the energy range of $\sim$~200~\keV\ to 20~\MeV,
with an angular resolution of about 3 degrees (1 sigma) at 2 MeV, within a
field-of-view (FOV) of about 1~sr. The detector's energy and three-dimensional spatial
resolution as measured during pre-flight calibration experiments, are $ \Delta
E_\mathrm{lxe}/E=8.8\% \: \sqrt{1\MeV /E}$ and $< $~1mm RMS, respectively.  The
detection efficiency for Compton events varies between 1.5 \% and 4 \% depending
on energy and event selection.  We describe the instrument as flown on May 7,
1999 and review its overall performance at balloon altitude. The launch occurred
at 13:26:54 UT from Ft.~Sumner, New Mexico and the flight was terminated about 9
hours later. The Crab was in the instrument FOV for a few hours. Analysis of
these data is in progress.
\end{abstract}


\section{Introduction}
In 1989 the concept of a liquid xenon time projection chamber (LXeTPC) as a
Compton telescope for MeV \g--ray astrophysics was proposed\cite{EAprile:89:SPIE}~. A series of experiments with LXe detectors followed,
aimed at establishing the charge and light yields, with specific emphasis on the
use of LXe in a position sensitive detector for MeV gamma-ray astrophysics
\cite{EAprile:89:IEEE,EAprile:91:performance}~. In 1993, 
the design and construction of a 10 liters LXeTPC was started, in collaboration with Waseda University. 
The goal was to
demonstrate the technology with a laboratory prototype of a size suitable to
verify the spectroscopy and Compton imaging response to \MeV\ \g --rays.
Details on the mechanical design of the TPC electrodes structure and its
cryostat for low temperature operation, xenon purification and cryogenics,
readout electronics for charge and light signals, and results from extensive
laboratory tests which established its successful operation as a radiation
detector are discussed in \cite{r:fxu}. Based on these results it was proposed,
in late 1995, to use the LXeTPC prototype as a balloon-borne \g --ray telescope,
to ultimately demonstrate its imaging capability on astrophysical
sources. Turning the laboratory detector into a flight instrument (LXeGRIT) has
required new developments mostly in three areas: a cryogenics flight system, a
readout electronics and data acquisition flight system, and an instrumentation
and control system. A parallel effort has involved modifications of the gondola
and veto shield systems of the University of New
Hampshire Directional Gamma-Ray Telescope (DGT) \cite{Dunphy:89:DGT}~, made available for LXeGRIT.  
A description of the flight instrument and its
performance during its May 1999 balloon flight follows. Details on the LXeTPC
detector and its readout electronics can be found in \cite{EAprile:98:electronics}~. Results on its spectral and imaging performance
from pre-flight calibration experiments are presented in a separate contribution
to these proceedings \cite{EAprile:2000:Spie}~.

\section{Instrument Description}

\begin{figure}
\centering
\includegraphics[bb=66 505 743 1044,width=0.8\textwidth,clip]{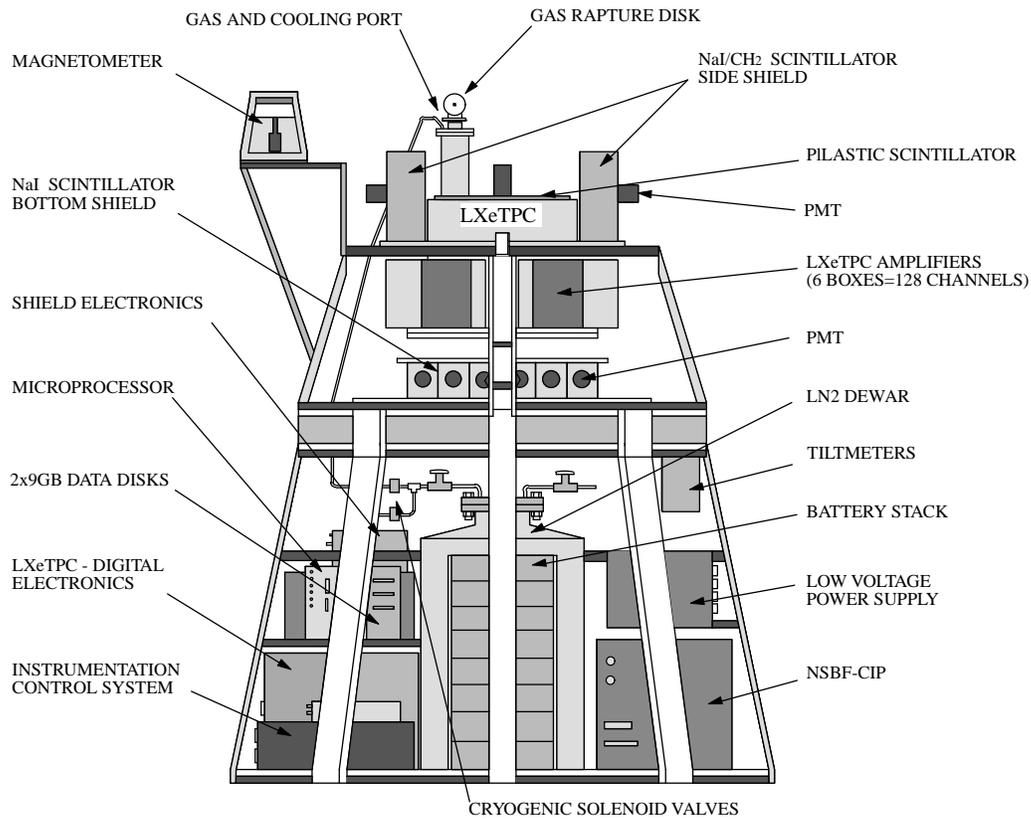}
\caption{Schematic of LXeGRIT payload in 1999 flight configuration.}
\label{f:gondola}
\end{figure}

\begin{figure}
\centering
\psfig{file=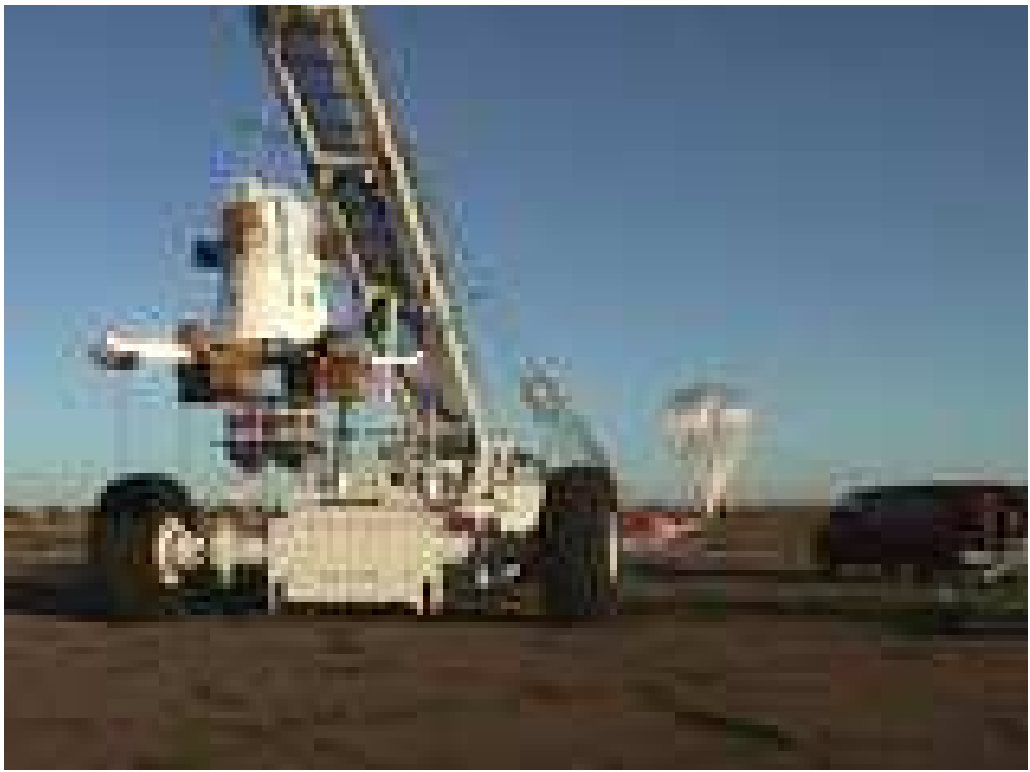,width=0.8\textwidth,clip=}
\caption{LXeGRIT on the launch pad on May 7, 1999}
\label{f:launch}
\end{figure}

A schematic of the LXeGRIT instrument assembled on the balloon gondola is shown
in Fig.~\ref{f:gondola}. The actual payload, suspended from the launch vehicle a
few hours before the flight is shown in Fig.~\ref{f:launch}. LXeGRIT, in its May 1999 flight configuration,
was a 1100 kg payload, exclusive of the balloon, parachute and ballast.  
Table~\ref{t:lxegrit} summarizes the instrument characteristics.

\begin{table}[htb]
\vspace{\baselineskip}
\centering \sloppy 
\abovecaptionskip 5pt
\tabcolsep 2pt
\begin{tabular}{|ll|}\hline
\tabcolsep 2pt
\raggedright
Energy Range & $0.2$ -- $20$ MeV\\
Energy Resolution $(FWHM)$  & $8.8\%\times(1\,\mathrm{MeV}/E)^{1/2}$ \\
Position Resolution $(1\sigma)$ & 1 mm (3 dimensions)\\
Angular Resolution $(1\sigma)$ & 3\deg\ at 1.8~\MeV \\
Field of View (FWHM) & 1~sr \\
Effective Area (Imaging) & 16~cm$^2$ @ 1~MeV \\
LXeTPC Active Volume & 20 cm $\times$ 20 cm $\times$ 7 cm \\
Active Back Shield & 2730 cm$^2$, 10 cm thick NaI(Tl) \\
Active Side Shield & 4750 cm$^2$, 10 cm thick NaI(Tl) \& CH$_2$ \\
Active Top Shield & 1600 cm$^2$, 1.2 cm thick plastic \\
LN$_2$ Dewar & 100 liter \\
Instrument Mass,\enspace Power & 1100 kg,\enspace 450 W \\
Telemetry,\enspace Onboard Storage & $2 \times 500$~kbps,\enspace 
                                     $2 \times 9$~GB \\
\hline
\end{tabular}
\caption{LXeGRIT Payload Characteristics.}
\label{t:lxegrit}
\end{table}

\begin{figure}
\centering
\includegraphics[bb=70 213 562 550,width=\textwidth,clip]{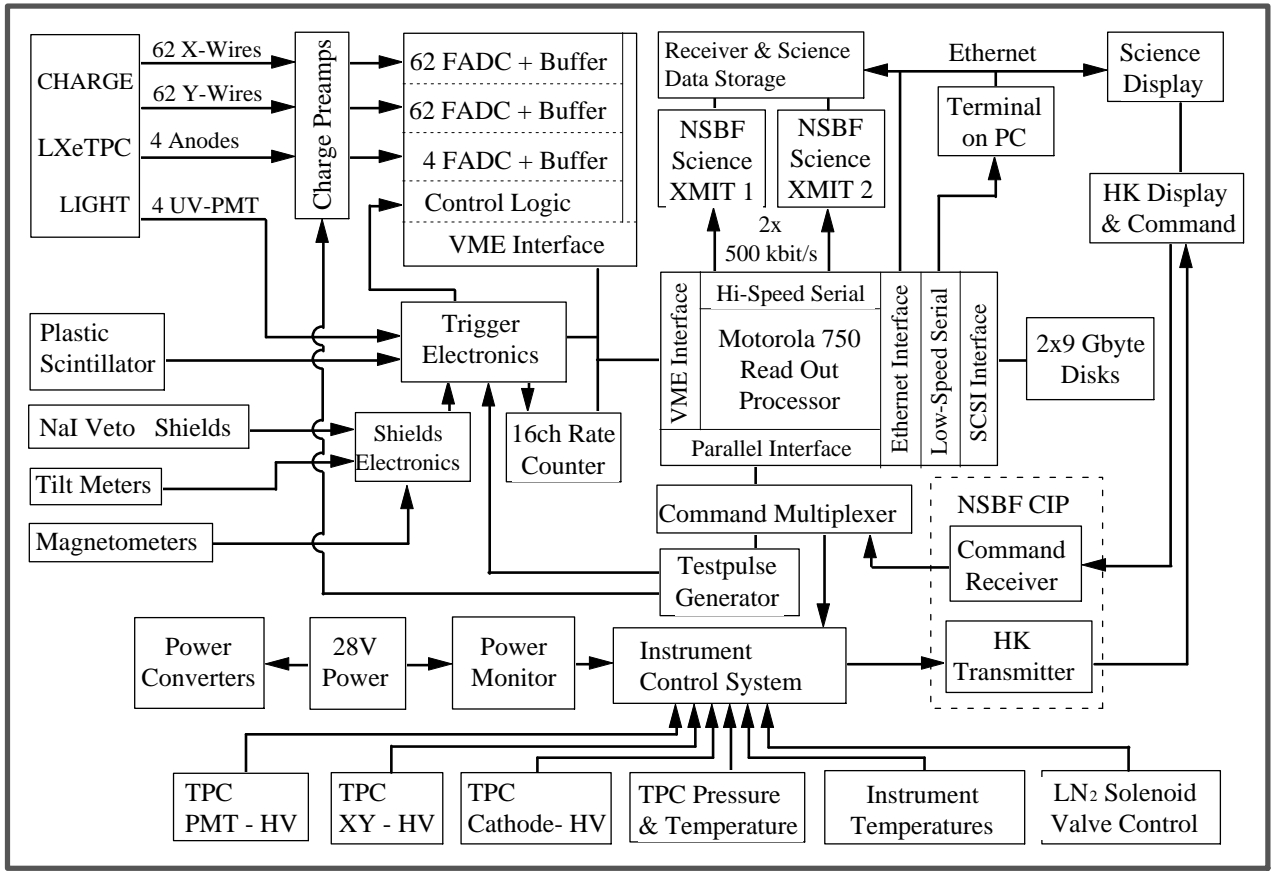}
\caption{Block diagram of the LXeGRIT system in the 1999 balloon flight. Also
indicated are additional interfaces such as terminal and Ethernet connection of
the DAQ processor for laboratory testing and calibration and the basic control
units on the ground.}
\label{f:fltsys}
\end{figure}

The DGT gondola was extensively modified for LXeGRIT. Since the instrument is a
Compton telescope, it does not require an active pointing system. The
instrument's zenith direction is determined from the pitch and roll with respect
to the local vertical, measured by two precision inclinometers. Azimuth
orientation is provided by two crossed magnetometers, calibrated before the
flight. This information, together with the knowledge of the geographical
coordinates of the payload, is needed for imaging analysis of the flight data. A
simplified block diagram of the LXeGRIT flight system is shown in
Fig.~\ref{f:fltsys}. We refer to both figures in the following description of
the various instrument components. Since LXeGRIT was built around a pre-existing
detector system, a pressure enclosure for a flight system would have been
difficult and expensive to realize. Therefore, special care had to be paid to
insure that every subsystem would sustain the near vacuum conditions and the
temperatures encountered during a balloon flight.
 
\subsection {The LXeTPC}

\begin{figure}[p]
\centering
\includegraphics[width=0.75\textwidth,clip]{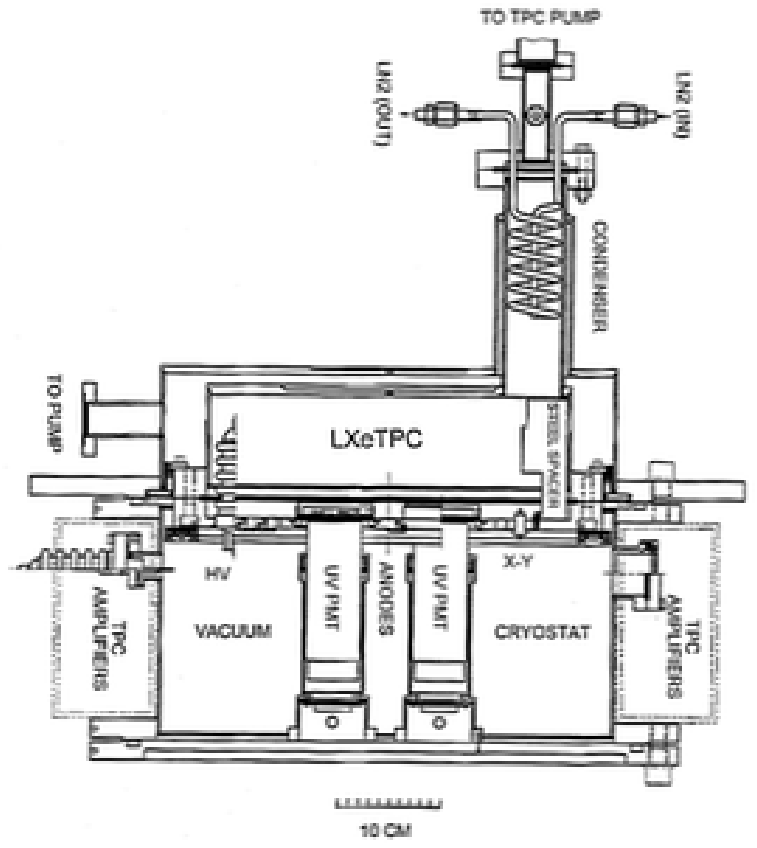}
\caption{\label{f:TPC} The LXeTPC mechanical design.}
\bigskip
\includegraphics[width=0.48\textwidth,clip]{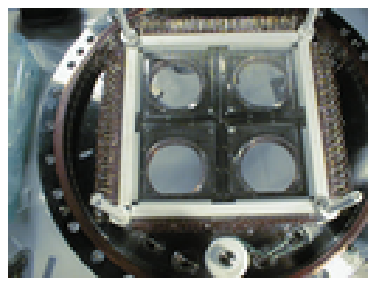}
\hfill
\includegraphics[width=0.48\textwidth,clip]{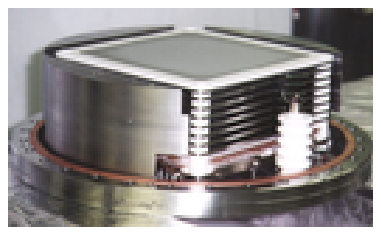} 
\caption{\label{f:TPC:images}
The TPC mounted on its bottom flange before and after complete assembly.}
\end{figure}

At the center of LXeGRIT is the LXeTPC. A simplified mechanical drawing is shown in Fig.~\ref{f:TPC}. 
The TPC electrodes are mounted on a 42 cm diameter conflat flange (see
Fig.~{\ref{f:TPC:images}) which encloses a 10 liter cylindrical vessel. The
sensitive volume is $20 \times 20 \times 7$~cm$^3$, defined by the cathode at a
distance of 7 cm from the shielding grid and by the field shaping rings. The 7 cm
layer of active LXe corresponds to 21 g/~cm$^2$ of absorber. The cylindrical
vessel is not the optimum choice for the TPC structure, as it leads to a large
amount of LXe outside the active volume. To reduce this amount, and thus
minimize the presence of an effective but un-vetoed scintillator around the
active region, spacers made out of solid steel are used, except on the cathode
HV feedthrough side. These spacers, clearly shown in Fig.~{\ref{f:TPC:images},
reduce the amount of LXe needed to fill the vessel by about 2~l.  Four quartz
windows are brazed on the detector's flange to couple four UV PMTs to the
chamber, for the Xe scintillation light detection. Hermetic feedthroughs for
signals (124 for the X-Y wires readout structure, 4 for the anodes) and HV lines
are also welded on this flange. High purity ($< 1$ ppb O$_2$ equivalent) Xe gas
is liquified into the vessel by a controlled flow of liquid nitrogen (LN$_2$)
through the copper coil of the condenser on the top. The liquid temperature is
$\sim -95\deg$~C, at a pressure of 1.5 atm. Thermal insulation of the cold
vessel is provided by a vacuum cryostat. The lower section of the cryostat
encloses the four PMTs, as well as the HV distribution circuitry for the wires
and the cathode. Both TPC vessel and vacuum cryostat are made out of stainless
steel. The outer dimensions of the complete detector are about 50 cm diameter by
35 cm high, excluding the condenser. The size of the lower cryostat is determined by the
length of the PMTs. The total mass, including the liquid xenon,
is about 190 kg.

\subsection {The Cryogenics System}

The liquid xenon, once in the detector, is maintained at the desired temperature
by a controlled flow of LN$_2$ through the condenser. During cooling,
microphonics noise affects the stability of the digitized pulses, mostly the
anode ones, and therefore the quality of the data. To minimize the amount of
unusable data, we allow the vapor pressure on top of the liquid to build from
1.4 atm to 2.4 atm. At this point, a solenoid valve opens to start the flow of
LN$_2$ and it closes once the pressure is down to 1.4 atm. Both the vapor
pressure above the liquid and the temperature of the bottom flange are monitored
continuously. We do not see an impact of the few degrees temperature variation
on the data. In the laboratory, a cooling cycle lasts less than 10 min and the
period of the cycle is about 75 min.  During the flight, the period is longer
due to the colder air temperature (-20 \deg\ at a float altitude of 40 km) and
the near vacuum conditions. For the cooling of the detector during the flight, a
LN$_2$ dewar is carried on board. It is a commercial dewar, modified to include
a molecular sieves trap, to maintain the vacuum in the TPC cryostat during the
flight. The hold time for the flight dewar with a capacity of about 90 liters,
is about 40 hours.  For redundancy, two solenoid valves are used on board,
controlled by the Instrumentation and Control system (ICS). In case the primary
valve fails to open, the secondary valve can be opened manually by ground
command. The system has worked reliably both in the last flight and in the 1997
test flights from Palestine, TX.  The TPC is equipped with a rapture disk, set
at a pressure of 7.5 atm. When the payload is released from the balloon and
starts the descent, the cooling system is kept powered. During the May 1999
flight, the payload was recovered a few hours after touch down, in a location
about 350 miles from Ft Sumner. After replenishing the LN$_2$ supply, the
cooling system kept working until the payload was brought back to the launch
base, more than 24 hours after recovery.  As the launch date and time is not
known a priori, the detector must be filled and operational well in advance (a
couple of days) of a launch opportunity. This requires a good purity of the
xenon, a low outgassing rate for the detector, and a reliable cooling
system. The TPC charge stability with time has been verified in laboratory
experiments lasting several days, as well as after the recovery of the payload
following the May flight.

\subsection{\label{s:veto} The Shield Veto System}
To reduce the charged particles and gamma-ray background rate at balloon altitude, it was decided to use the available DGT shields to cover the LXeTPC as much as possible. The detector intrinsic 3D event imaging capability allows an effective identification and rejection of single interaction events and charged particles in the off line analysis. However, the decision to use an active shield was motivated by the desire to reduce the overall data rate to be transmitted via telemetry to ground. 
The shield configuration used for the LXeGRIT balloon flights is shown in Fig.~\ref{f:veto}.
\begin{figure}
\centering
\includegraphics[bb=211 167 714 806,height=0.7\textheight,clip]{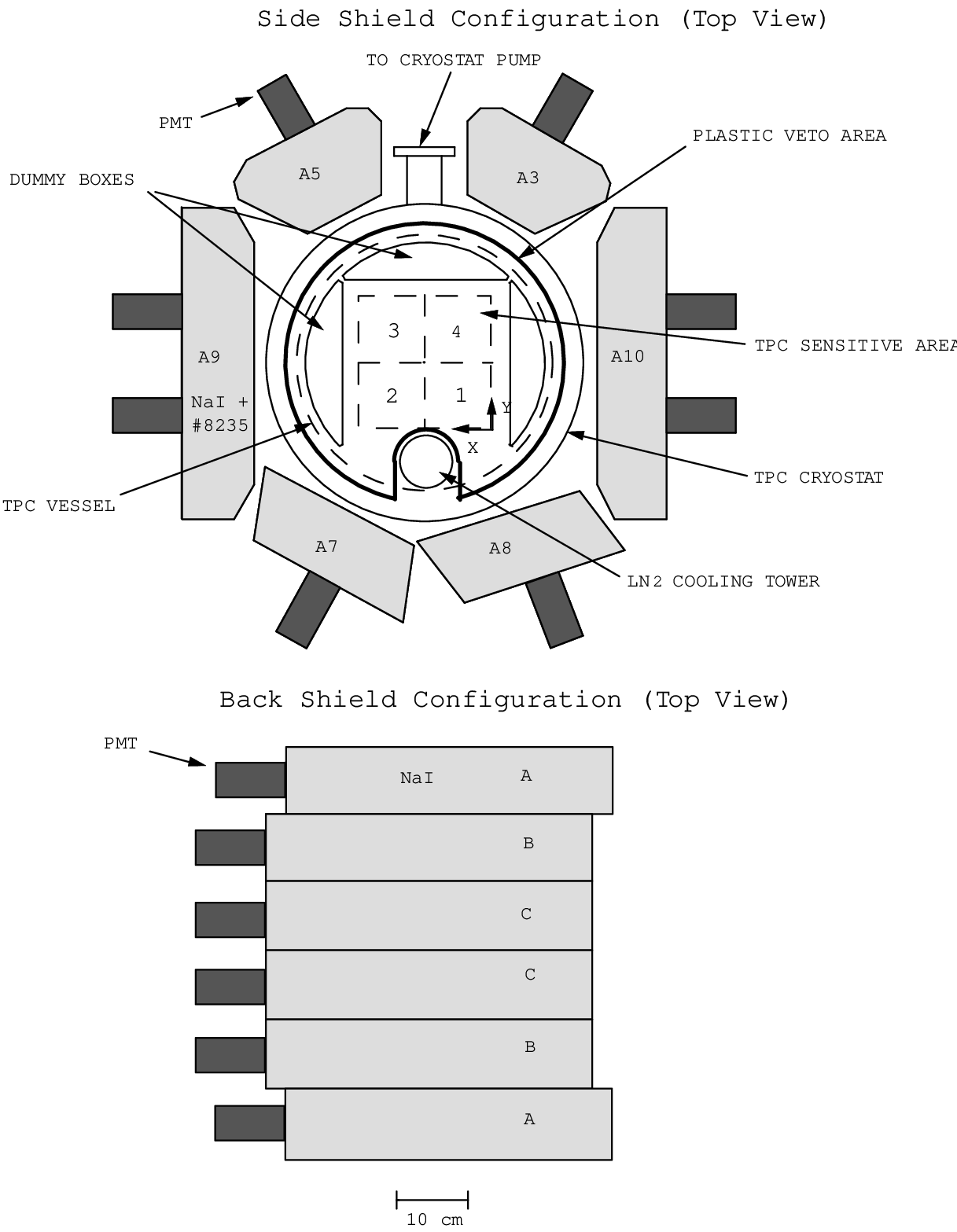}
\caption{Schematic of the shield configuration.}
\label{f:veto}
\end{figure}
The back shield consists of six 10 x 10 x 41 cm$^3$ bars of NaI(Tl) with PMTs at
one end, covering an area of about $2730$~cm$^2$. The side shield consists of
six units, each composed of small chunks of NaI(Tl), immersed in liquid
scintillator (NE235), sealed in an aluminum container. Each unit is viewed by
two PMTs coupled directly to the NE235 mineral oil. The units are of three
different shapes, but they all have a thickness of about 10 cm and a height of
25 cm, limiting the TPC field-of-view (FOV) to about 45 degrees. The packing
fraction of NaI in the side shields is about 75\%. The total mass of the shield
system, including the support structures electronics and cables, is about 270
kg.  The energy threshold of the side shields, which have poor energy
resolution, was adjusted to match the rate in the NaI back shields with a
threshold around 300 keV.  The original DGT shield electronics and control logic
was used. The individual shield rates together with the attitude information
from the magnetometers and inclinometers were downlinked to ground through a
separate telemetry channel.  A plastic scintillator counter of 1.2 cm thickness
and an area of $1600$~cm$^2$ was added to veto the charged particles entering
the TPC from the top. It is mounted directly on top of the TPC cryostat and is
viewed by one PMT in the center. Its signal is measured by one channel of the
UNH shields electronics. 
The logical OR of the signals from all the shield sections, including
the plastic, is fed into the LXeGRIT Trigger Electronics System, to veto the LXeTPC light trigger signal, as discussed below.

\subsection {The Charge and Light Readout Electronics System}
  
The LXeTPC detects both ionization and scintillation light signals, produced by gamma-ray interactions or by charged particles. Free electrons are drifted in a uniform electric
field of 1~kV/cm for a maximum distance of 7 cm, corresponding to a maximum drift time
of 35 $\mu$s. The X-Y position of an ionizing event is determined from the
signals induced on two orthogonal planes of parallel wires with a 3~mm
pitch. The Z-coordinate, is determined from the drift time, measured with
respect to the trigger of the fast ($<5$~ns) Xe light. The energy deposition is
measured on four anodes each covering an area of about $10 \times 10$~cm$^2$.
Each of the 62 X-wires and 62 Y-wires and each anode is amplified with a charge
sensitive amplifier and fed to a custom-built digital electronics system. The
signals are digitized at a rate of 5~MHz with a resolution of 8-bits for the wires
and 10-bits for the anodes. The digitized data are stored in an array of ring
buffers with 256 samples per event, covering more than the maximum drift time. The maximum energy which can be
sensed by this system is around 20~\MeV. The design and fabrication of the
LXeGRIT digital electronics was provided by Marshall Space Flight Center and is
described in \cite{EAprile:98:electronics}. Low-power consumption, while
desirable, was not a primary concern in the design of this system. The average
power consumption of the tightly packed digital electronics system accounts for
about half of the total LXeGRIT power.  As discussed in reference
\cite{EAprile:98:electronics}, problems of overheating were encountered during the first LXeGRIT test flight of a few hours from Palestine, TX in July 1997. These problems were corrected for the 1999 flight, by better passive distribution from the heat source to the gondola frame.
The frontend charge sensitive amplifiers system, also described in reference
\cite{EAprile:98:electronics}, are housed in six boxes mounted around the lower
section of the cryostat.  The system is such that the equivalent noise charge on
the wires is typically less than $\sim$~400~e$^-$~RMS, while the noise on the
anodes of higher capacitance is $\sim$~800~e$^-$~RMS. With these noise
conditions, the TPC can well detect the multiple interactions of \MeV \
\g--rays, with energy deposits as low as $\sim$100~\keV, in optimized
conditions.

The trigger for the digital electronics is provided by a light signal from the
TPC, if not vetoed by a signal in the shields or the plastic counter.  After
amplification, the signals from the four PMTs viewing the
LXe volume, are passed through four window discriminators. The lower-level
discriminator reduces noise pulses, whereas the upper-level discriminator is
used to discriminate against high energy events, mostly from charged particles
traversing the TPC. These thresholds are adjustable in flight, via commands
from the data acquisition processor. The upper threshold can also be turned
off. The OR of the four PMTs constitutes the TPC light trigger. In case of
no veto signal from the shield system, the TPC light trigger starts the data acquisition and
marks the beginning of the drift time measurement. To avoid readout of more than
one event in the sensitive volume, it is required that there was no PMT signal
within 50 $\mu$s before the event. If a second PMT signal occurs within 50 $\mu$s after the original trigger, an abort signal is generated to stop data acquisition and to reset the system.

\subsection {The Data Acquisition System}

Data acquisition (DAQ) is controlled by a commercial processor (Motorola
MVME2700) with good computational performance (10 specInts), high bandwidth
architecture ($>100$~Mbytes/s on the internal PCI bus) and all the I/O
subsystems (parallel port, high speed serial ports, SCSI bus) needed in the DAQ
layout. The new processor board, placed in a separate pressurized box, is
interfaced to the trigger electronics and the waveform digitizers via a VME-like bus,
using a special adapter card to interconnect with the custom-made bus of the
existing digital electronics. The microprocessor polls the digital front-end
electronics for data and reads the digitized waveforms with a through-put of 400
-- 600~Kbytes/s. As soon as all signals are read out, it readies the electronics
for the next trigger. About a quarter of the accepted events are transmitted to
ground via two high-speed serial links at 500~kbits/s each, to provide feed-back
on the data quality. The remaining events are written via SCSI interface to one
of two onboard hard disks with 9~Gbytes capacity each. For a good understanding
of event patterns at float altitude in this new type of detector, data were
predominantly taken in a mode that reads all information from anodes and
wires. While providing a full image of the chamber continuously, this mode
requires an event size of about 29~Kbytes, slowing down the DAQ to a rate of
typically 100 -- 150 handled events/s. About 7\% -- 12\% of these events were
accepted by the online checks performed by software:
\begin{itemize}
\item Requirement of a minimum signal amplitude above initial baseline on at
      least one anode. 
\item Requirement of a minimum and a maximum number of wire hits on each view 
      (X and Y), i.e., wire signals above thresholds recognized in hardware.
      Thresholds are either set manually or automatically by noise evaluation 
      with test triggers.
\end{itemize}

Data taking is controlled from ground by transmission of commands to a
multiplexer, which forwards DAQ commands to the microprocessor via a standard
parallel port. DAQ can thus be controlled by, e.g., setting thresholds in the
charge and light read-out and by changing the DAQ mode. Commands as well as
configuration packets which summarize DAQ settings are transmitted to ground via
the two fast ``science data'' channels, and are also written to disk to cope with the high data rate.
In addition, the DAQ processor reads the sixteen 16-bit counters in
the trigger electronics system, which record, e.g., the rate of the
individual PMTs, their logical OR within a given discriminator window and above
the lower level threshold. These rates serve as monitor of the detector
operation, and also determine the influence of each cut in the trigger decision
to finally calculate the flux of gamma-rays.  The trigger rates, together with
internal counters that keep track of the DAQ efficiency, are transmitted to ground every
two seconds. All the relevant information is displayed online.

\subsection {The Instrumentation and Control System}

Apart from the trigger rates, all the housekeeping and instrument control is
handled by a designated processor part of the Instrumentation and Control System. 
The control functions include setting the high voltages for the TPC cathode, X-Y wire planes
and PMTs, and to operate the solenoid valves for cooling of the 
detector. The corresponding commands are decoded from the signals 
received by telemetry. The high voltage on the electrodes is raised and lowered at a 
predetermined rate to avoid an excessive charge build up due to the 
voltage change, which could destroy the input of the charge sensitive 
amplifiers.
During the flight the status of the TPC and all the support systems 
are constantly monitored. The parameters measured by the ICS are the 
battery voltage and current, all high voltage values, the pressure in 
the TPC, and the reading of 16 temperature sensors 
distributed over strategically chosen locations on the payload. The 
temperatures give a crucial information. Although the ambient 
temperature at float altitude is typically -20 \deg, the reduced air pressure results in 
a heat transfer that is 20 times lower than on the ground. Any electronic
device might overheat if the produced heat is not conducted away sufficiently fast. Additional to these 16 temperature sensors, also the 
temperature of the liquid Xe is constantly measured. Because of the 
low temperature and of the precision required for this measurement, a 
platinum resistor mounted in good heat contact with the 
bottom flange of the TPC vessel is employed. Given the relation 
between vapor pressure and temperature, this measurement can be used 
to manually initiate cooling if the pressure transducer should 
fail shutting down the automatic system.

The operation of a HV system is problematic at reduced air pressure. 
With the reduced density, discharges are frequent at voltages 
completely safe at normal air pressure. These discharges of course 
dissappear in a good vacuum. The residual pressure at float altitude is about 
the worst condition for voltage breakdown. The cathode voltage power supply is therefore 
mounted within the envelope of the TPC cryostat which is evacuated 
for thermal insulation of the cold detector. The vacuum in the 
cryostat, provided by the cryo pump within the LN$_2$ dewar, is 
also measured by the ICS.

Whereas the applied voltage is a good measure to 
determine the field distributions in the TPC, it is the current in 
the divider chain which tells if all electrodes are supplied with the 
proper potential. Therefore also the currents in the divider chains 
for the cathode and field shaping rings, and in the X-Y wire planes HV 
distribution are constantly measured and transfered to ground by the 
ICS.

For the 1999 flight, primary low-voltage (28V) electrical power was supplied by a stack of 20 lithium batteries, rated at 30Ah each, allowing a flight duration of about 40 hours. 

\section{Instrument Performance}

LXeGRIT was launched from the National Scientific Balloon Facility (NSBF) in
Ft.~Sumner, NM, on May 7, 1999 at 7:26:54 local time (13:26:54 UT), using a 28
million cubic feet balloon. It reached float altitude after about 2.5 hours, and
remained there for 7~hours, before flight termination due to wind conditions.
During the entire flight including ascent, a total of 285966 events were
collected. The main goal of the flight was a measurement of the background rate
in the instrument and to verify the background discrimination capability of the
LXeTPC. To verify the Compton imaging performance, the launch time was scheduled
such as to have the Crab nebula in the FOV of the instrument for a few hours,
once the payload reached float altitude.  Fig.~\ref{f:fltprof} shows the
altitude of the payload during the flight. The atmospheric depth varied between
3.7 and 5.2 g/$cm^2$. LXeGRIT performed well during the flight: the LXeTPC and
its high voltage and cryogenics systems, the electronics and data acquisition
systems worked in the near space environment as they did in the laboratory.  A
few problems were encountered with some of the side shield units. The loss of
telemetry contact with the ICS for a major part of the flight did not impact the
science data.
The payload was recovered in good conditions, with mechanical
damages to the magnetometers and three side shield units. 
\begin{figure}
\centering
\includegraphics[bb=70 405 547 
663,width=0.7\textwidth,clip]{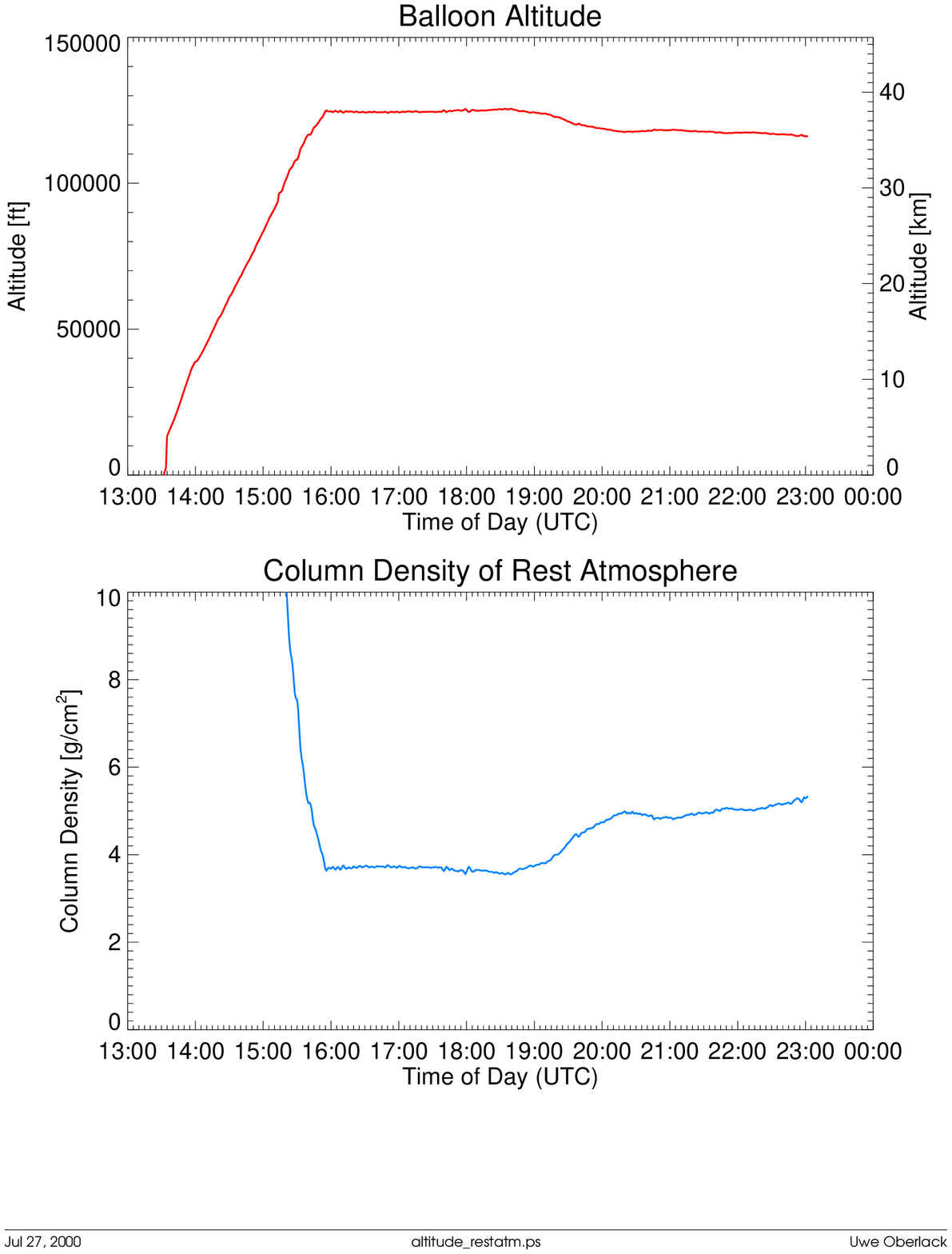}
\caption{\label{f:fltprof}
  Altitude of the balloon payload during the May 1999 flight.}
\end{figure}

\subsection {Preliminary Results from Flight Data}

\begin{figure}
\centering
\includegraphics[bb=70 485 513 729,width=0.49\textwidth,clip=]{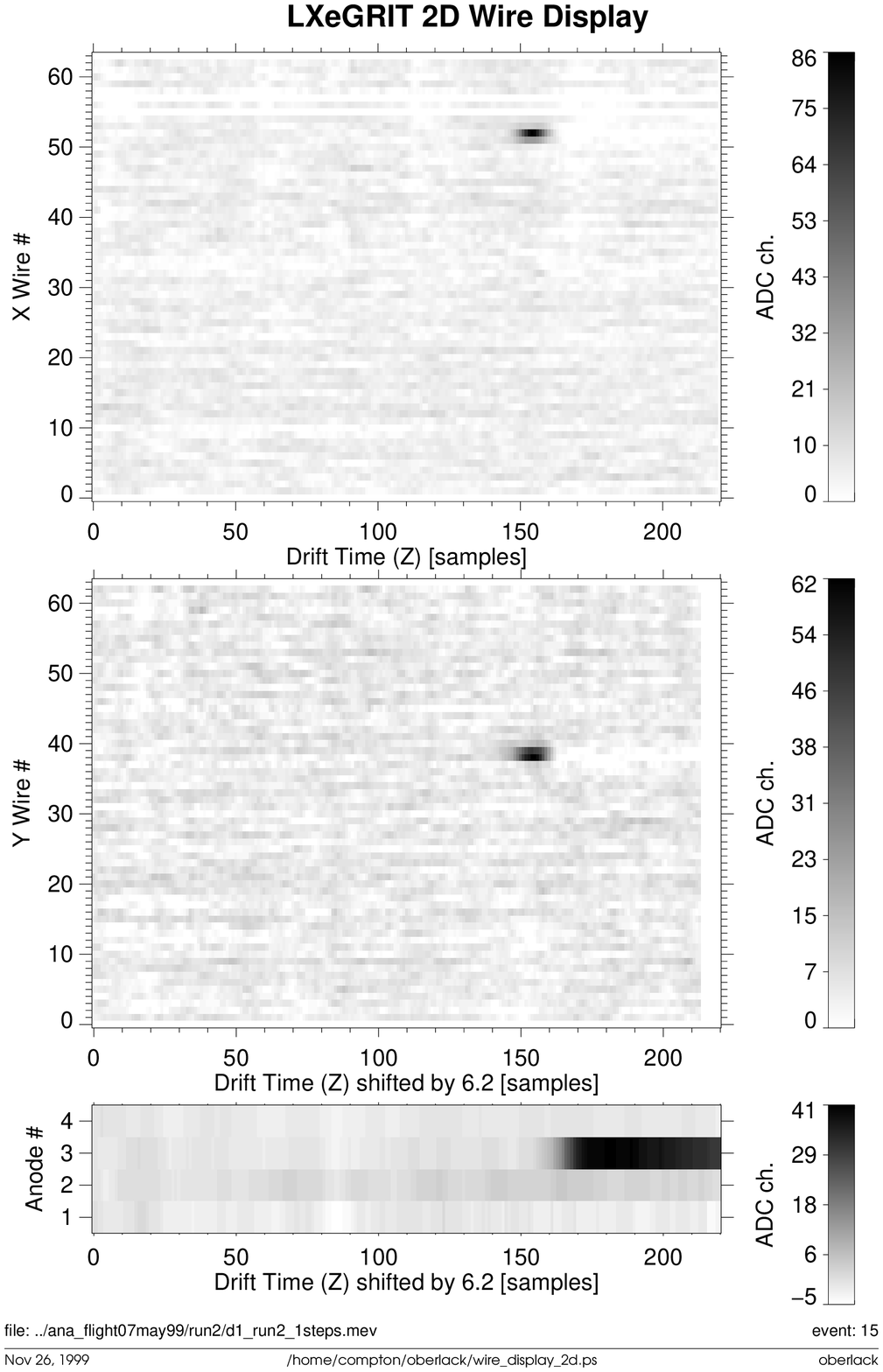} 
\hfill
\includegraphics[bb=70 485 513 729,width=0.49\textwidth,clip=]{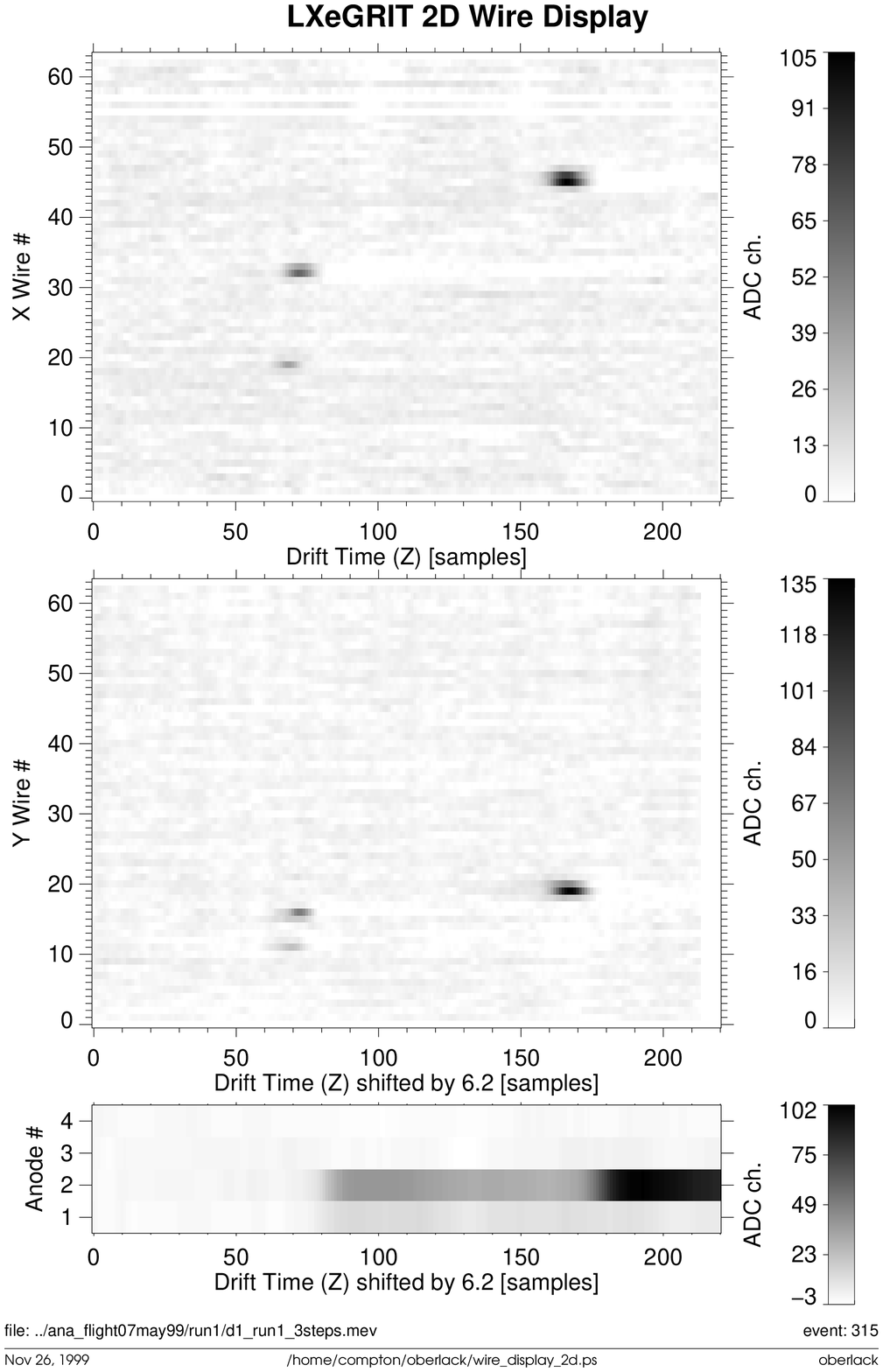} 
\\
\includegraphics[bb=70 454 513 729,width=0.49\textwidth,clip=]{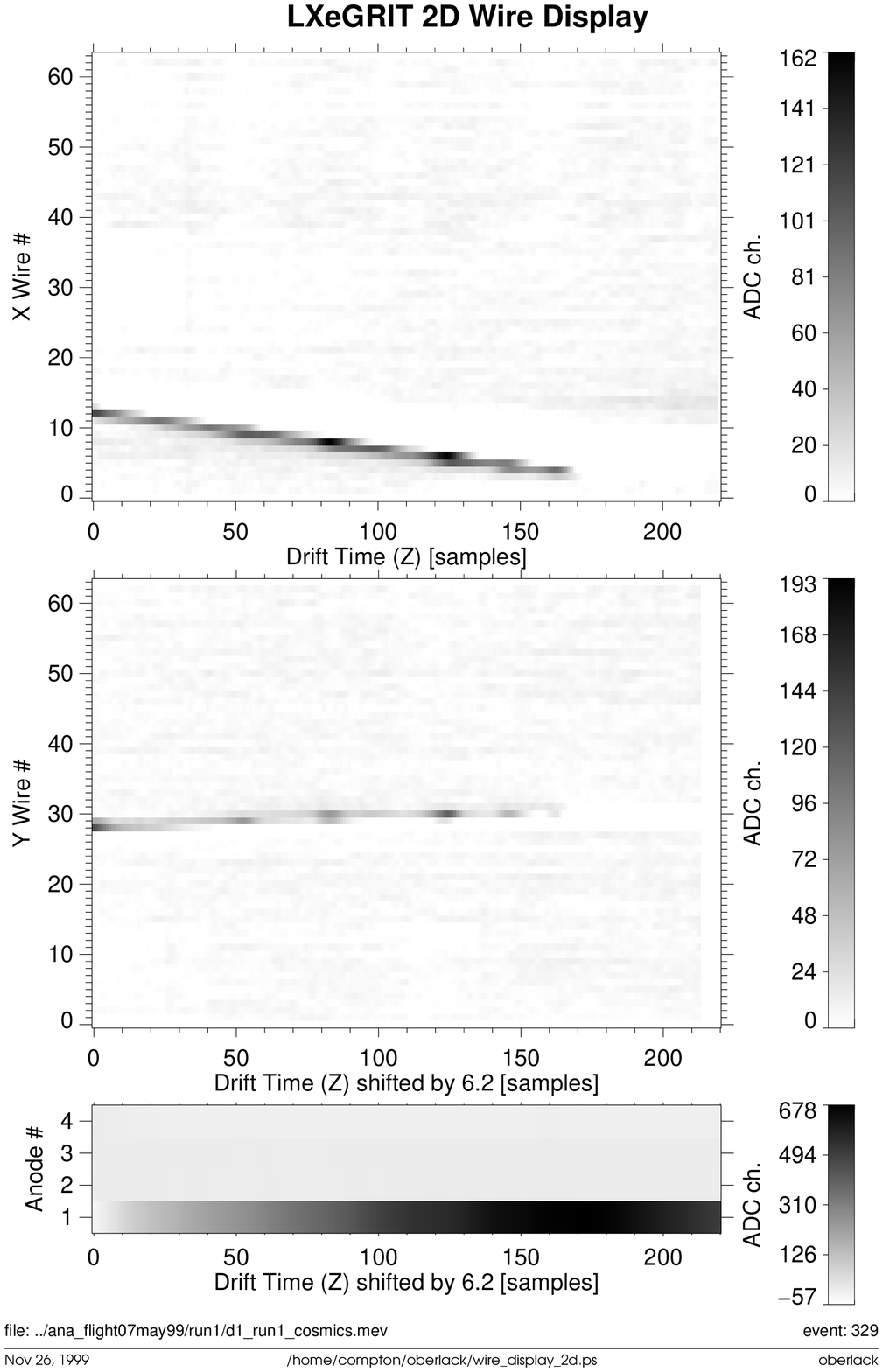} 
\hfill
\includegraphics[bb=70 454 513 729,width=0.49\textwidth,clip=]{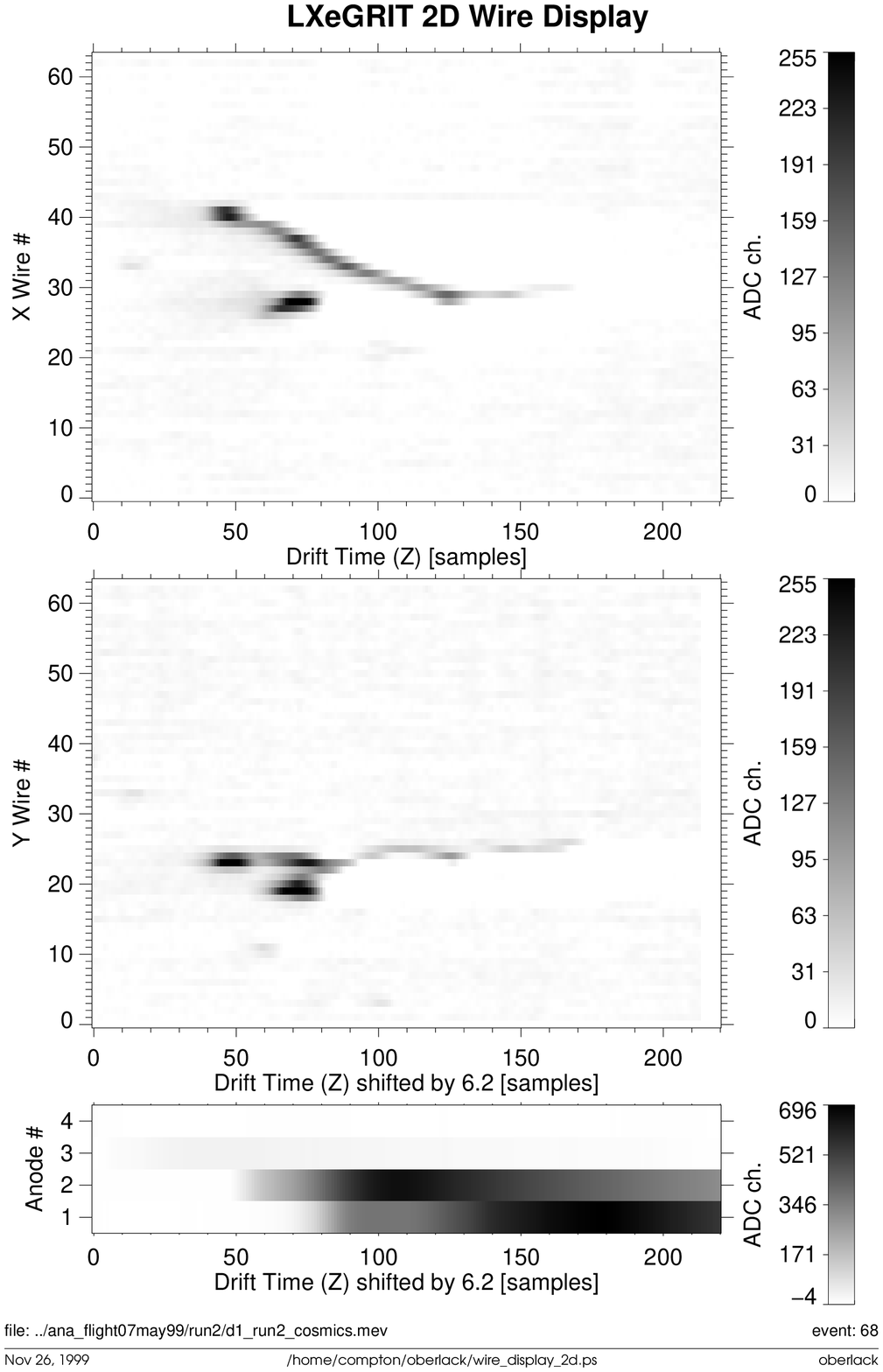}
\caption{\label{f:event-images}
 Different event types are easily recognized by their signature on the
 sensing wires, as shown in these images of ($X,Z$)-coordinates. Single- (top
 left panel) and multiple-interaction (top right panel) \g-events, a
 cosmic-ray track (bottom left), and a more complex event topology
 involving inelastic scattering (bottom right).}
\end{figure}
Fig.~\ref{f:event-images} demonstrates how 3D event imaging makes it possible to
distinguish between photons that undergo multiple Compton interactions, which
can be used for source imaging, from those that are absorbed in a single
interaction or within a small volume of the detector.  This provides very
significant background rejection. Only photons that produce multiple Compton scatterings
in a volume on the order of the spatial resolution of the detector are confused
with single-interaction events and therefore lost for imaging.  Charged
cosmic-ray particles are also easily recognized, both from their large energy
loss and long track length. Other event topologies, easily distinguishable from
Compton \g-events, result from fast neutrons generating nuclear reactions within
the detector volume.

\begin{figure}
\centering
\includegraphics[bb=59 38 564 712,width=0.7\textwidth,clip]{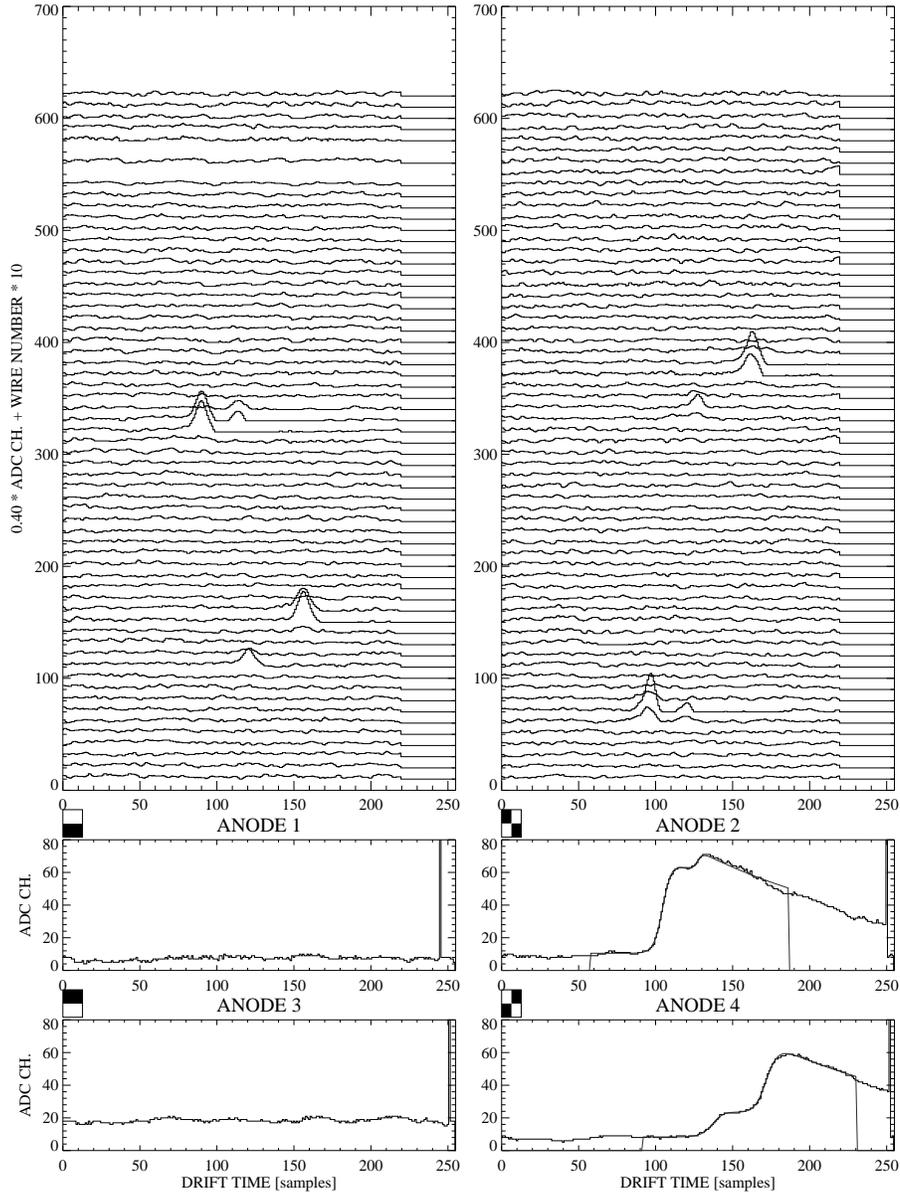}
\caption{\label{f:3step-evt}
 Display of a \g-ray flight event undergoing four interactions in the TPC. The noise
 conditions are comparable to those in the laboratory.}
\end{figure}

The noise conditions on anodes and wires during the flight were comparable to
laboratory conditions.  
Fig.~\ref{f:3step-evt} shows the complete display of the digitized wire and
anode waveforms as a function of drift time for a \g--ray event recorded during
the flight. From the signature on the anodes, where a fit of the anode waveforms
 is superimposed, the incoming photon is clearly recognized as making four
distinct interactions in the sensitive volume. The corresponding interaction
locations are also apparent on the X-Y wires. 

\begin{figure}
\centering
\includegraphics[bb=70 511 299 718,clip]{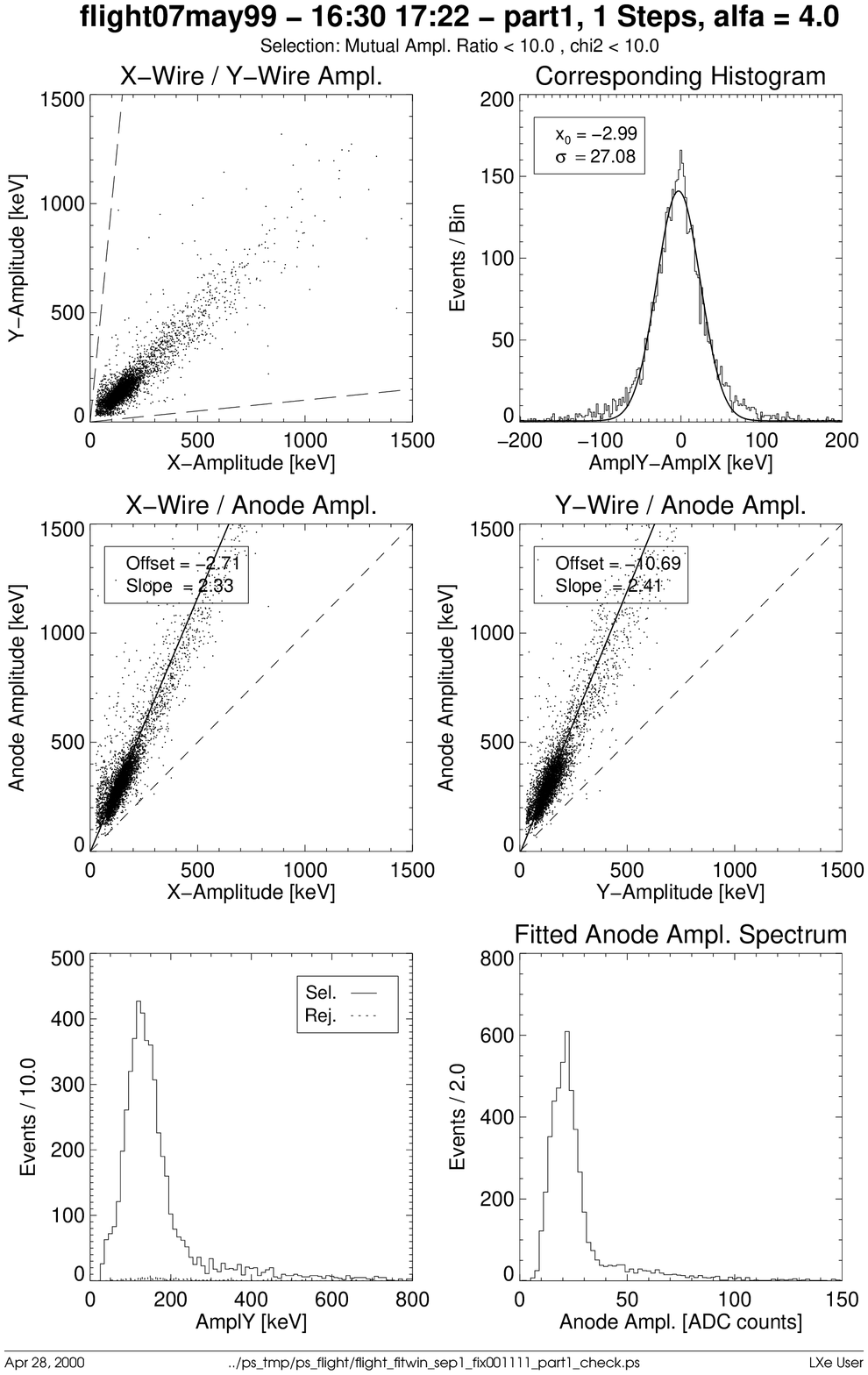}
\includegraphics[bb=295 286 525 491,clip]{match_1steps_p1.ps}
\caption{\label{f:match:amp}
 \emph{Left:} Correlation of signal amplitudes on X- and Y-wires.
 \emph{Right:} Correlation of signal amplitudes on anodes and Y-wires.
}
\end{figure}
\begin{figure}
\centering
\includegraphics[bb=70 511 299 718,clip]{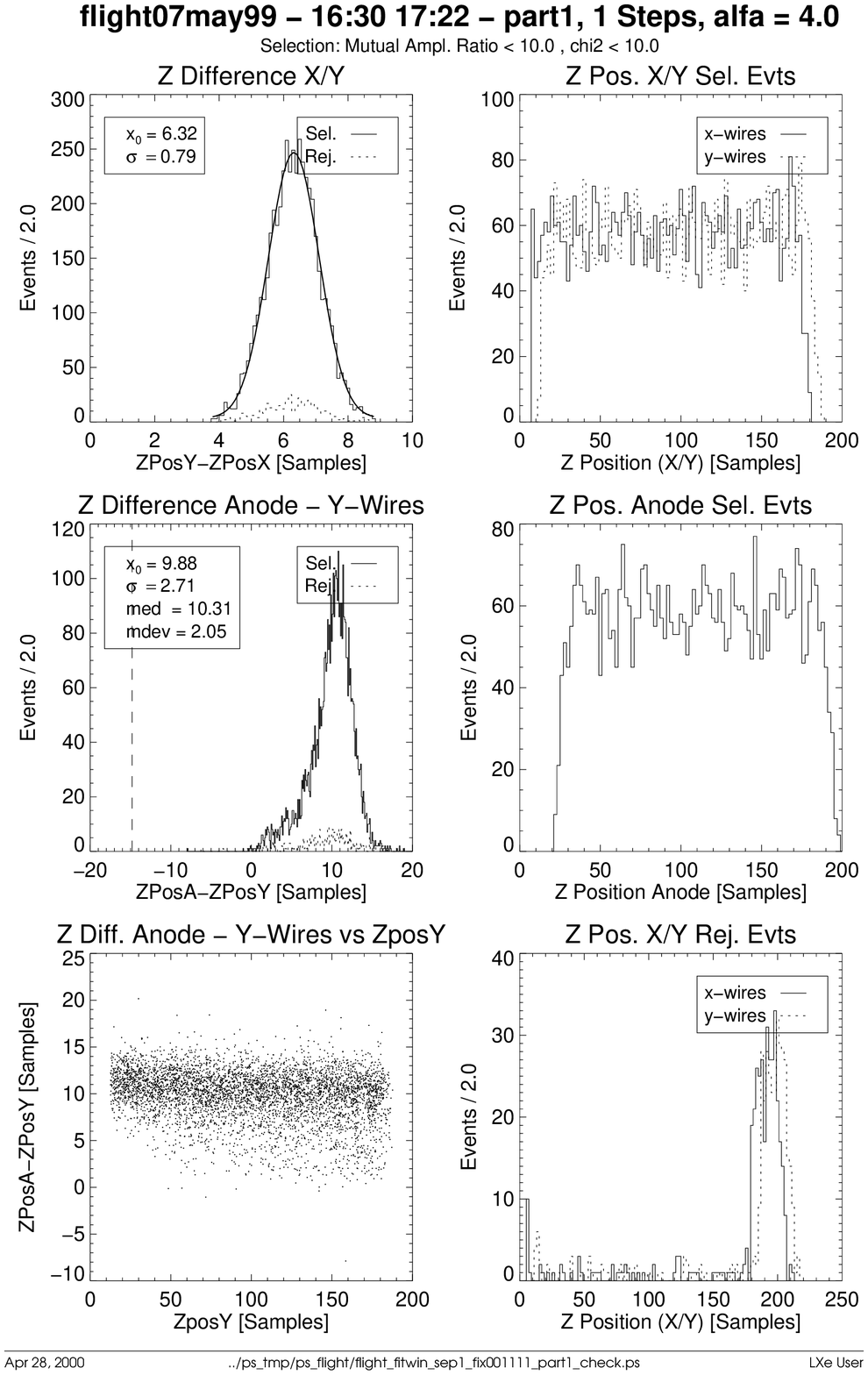}
\includegraphics[bb=81 511 288 718,clip]{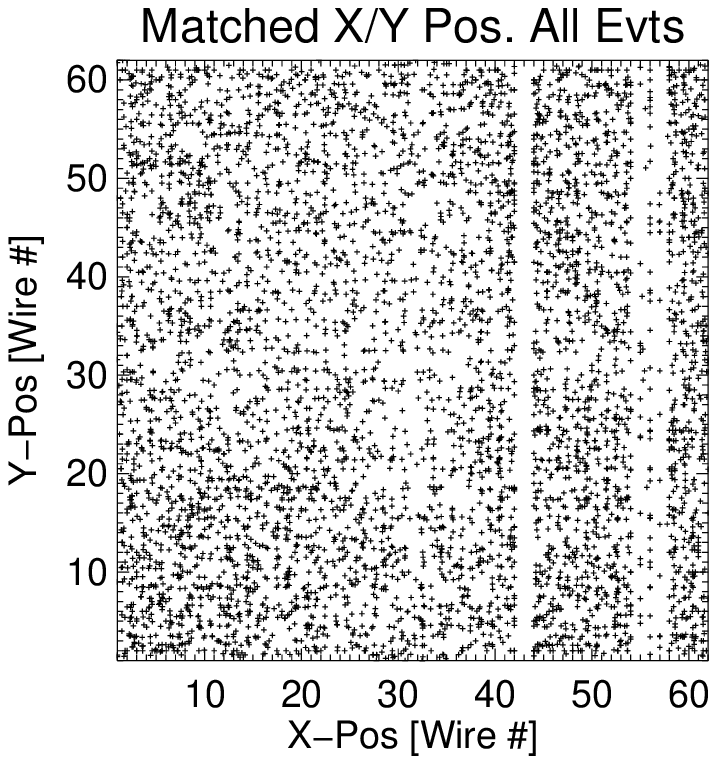}
\caption{\label{f:match:xy}
 \emph{Left:} Distribution of the drift time difference between X- and Y-wires,
 showing the same Z-accuracy as in the calibration.
 \emph{Right:} X-Y-distribution of single-interaction events.
}
\end{figure}

In the analysis of LXeGRIT data, the signals on anodes and wires are matched
according to drift time. The induction signals on both wire planes provide
redundant, even though less accurate, energy information in addition to the
pulse height measured on the anodes. Only a fraction of the current produced by
the drifting electrons is induced on the sensing X-Y wires. In the LXeTPC, the
amplitude of the charge signal induced on the wires is on average about 40 \% of
the total charge collected on the anode(s). The proper working of the signal
detection and matching algorithms can therefore be verified by comparison of X-
and Y-wire amplitudes, and their correlation with the anode amplitudes.
Fig.~\ref{f:match:amp} displays the amplitudes determined by the analysis
software for matched single-step events, which show a similar fraction of
induced charge 
versus collected charge as for calibration data (right panel in
Fig.~\ref{f:match:amp}). The precise matching of X- and Y-wire signals in drift
time is evident from the left panel in Fig.~\ref{f:match:xy}. The width and mean
of this distribution are essentially identical to the values found in
calibration data. With a digitizing rate of 0.2 $\mu$s per sample, the RMS value
of 0.79 samples corresponds to a timing accuracy of 160 ns, or an accuracy of
about 300 $\mu$m on the Z-position measurement.  The right panel in
Fig~\ref{f:match:xy} shows the power of an imaging detector, which allows us to
study the event distribution within the sensitive volume.  Single-interaction
events are predominantly found towards the outer rim of the detector.

A preliminary result for the in-flight background spectrum is shown in
Fig.~\ref{f:fltspec} from the analysis of about 3 hours of data. These are count
rates before deadtime correction. The upper figure is from all event types,
while the lower two figures show the contribution to the count rate from events
reconstructed with a single interaction (1-step) and with multiple interactions,
respectively. A very limited number of selections were applied to the data in
this figure. The analysis of these events and of the remaining event statistics
continues to-date, along with Monte Carlo simulations of all possible
contributions to the measured spectrum.

\begin{figure}
\centering
\includegraphics[bb=69 72 530 746,height=0.64\textheight,clip]{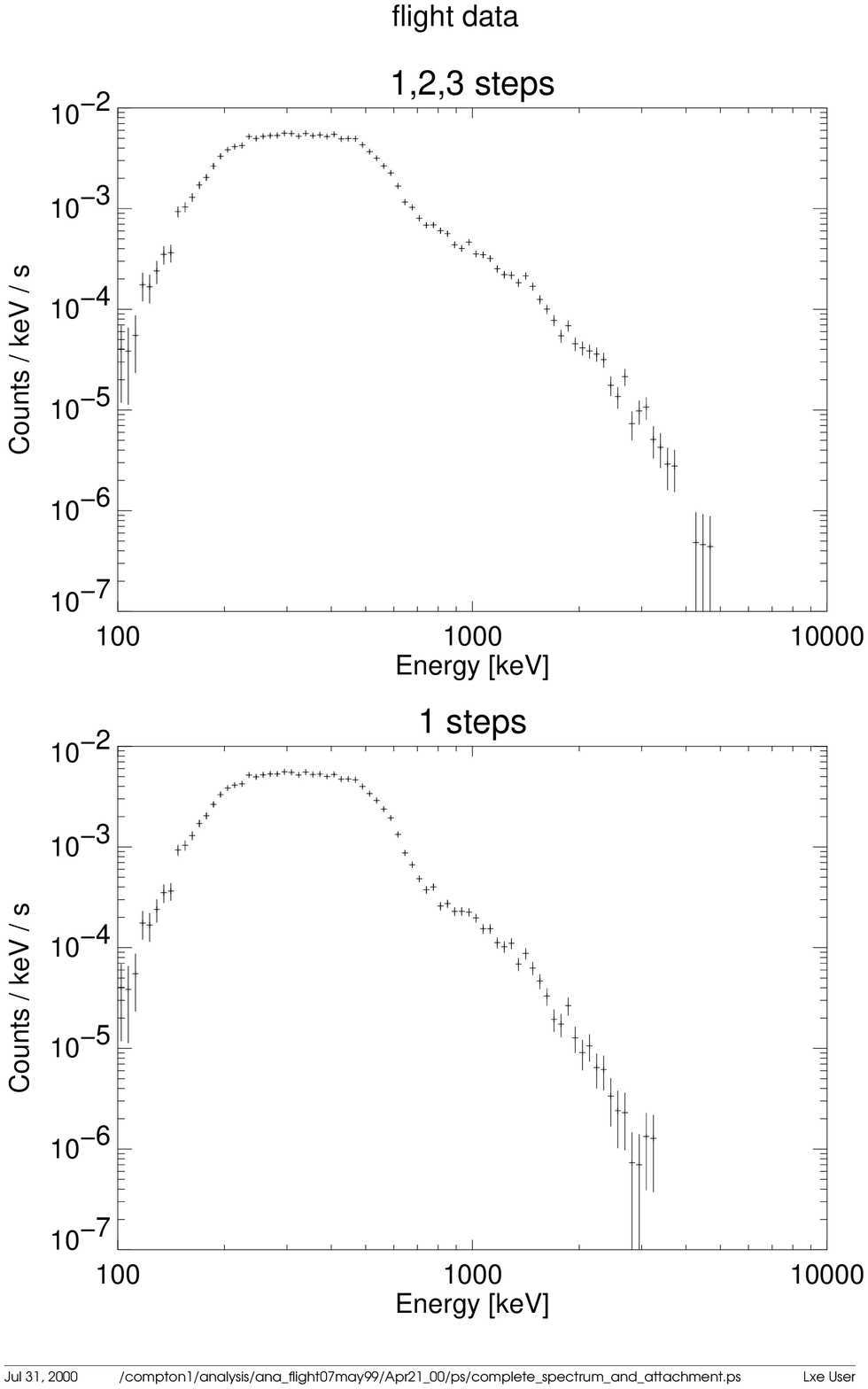} \\
\includegraphics[bb=69 414 530 746,height=0.31\textheight,clip]{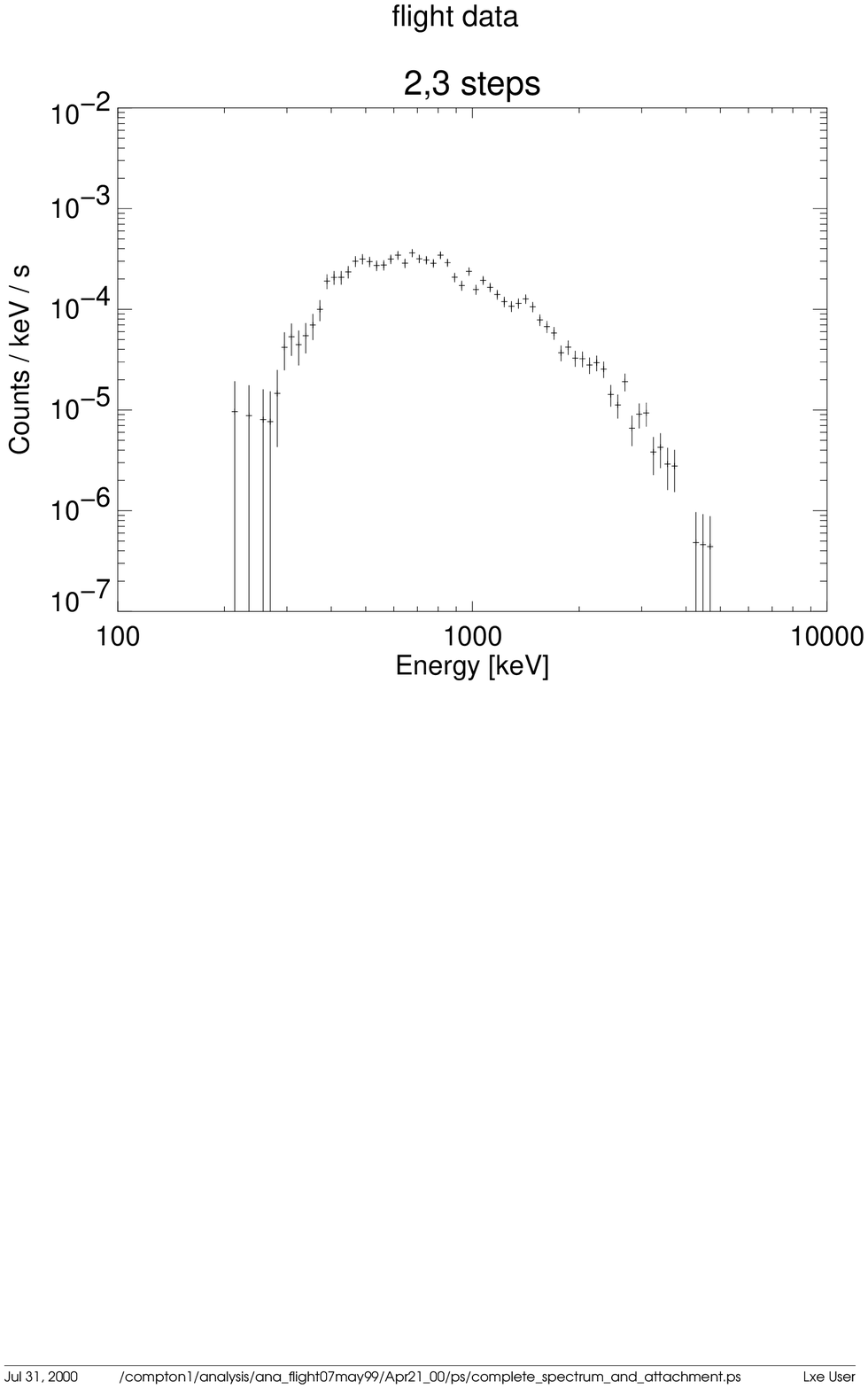}
\caption{Count rate spectra from about 3~h at float altitude during the 1999
balloon flight.
\emph{Top:}Events with 1 -- 3 interactions identified in the TPC and with 
matched signals on X- and Y-wires as well as on the anodes. 
\emph{Center:} Corresponding spectrum for single interactions only. 
\emph{Bottom:} Corresponding spectrum for 2 -- 3 interactions only. 
}
\label{f:fltspec}
\end{figure}
The spectrum is dominated by single interaction events. This type of events,
easily recognized by the TPC, are rejected for Compton imaging. It is however
intriguing that we see such a large number. Events with energy larger than a few
MeV are much less than we expected.  In this limited statistics sample, there is
a hint of a 511 keV line, as well as a line around 1.4 MeV.  We understand the
suppression of events with energy larger than a few MeV as due, in large part,
to the trigger selections used for the flight data taking.  The discriminator
window and the PMT high voltages were optimized to achieve a homogeneous trigger
efficiency throughout the detector volume, and to reject high energy charged
particles. This choice of settings coupled with the predominance of low energy
\g--rays background encountered at float, resulted in an enhanced rate of very low energy events, well below our range of interest.

A clear demonstration of the impact of this trigger selection on the measured event energy distribution is shown in Fig.~\ref{f:AmBe:spec}, from  laboratory data obtained with an AmBe source which emits 4.43 MeV \g--rays, along with neutrons. 
The energy spectrum on the left is from multiple interaction events recorded in the LXeTPC triggered with the same
window discriminator selection and the same HV settings on the four PMTs as in the flight. In this spectrum, there are almost no events in  the full energy peak at 4.43 MeV and in the first escape peak, expected from pair production which becomes increasingly important above 4 MeV. On the other hand, events at high energies dominate the energy spectrum on the right, obtained from an experiment with the same source but with a different trigger selection.  Clearly, the flight trigger selection enhances the low-energy part of the spectrum, largely reducing the efficiency for \MeV\ energy depositions. 
 
\begin{figure}
\centering
\includegraphics[bb=310 526 522 732,clip]{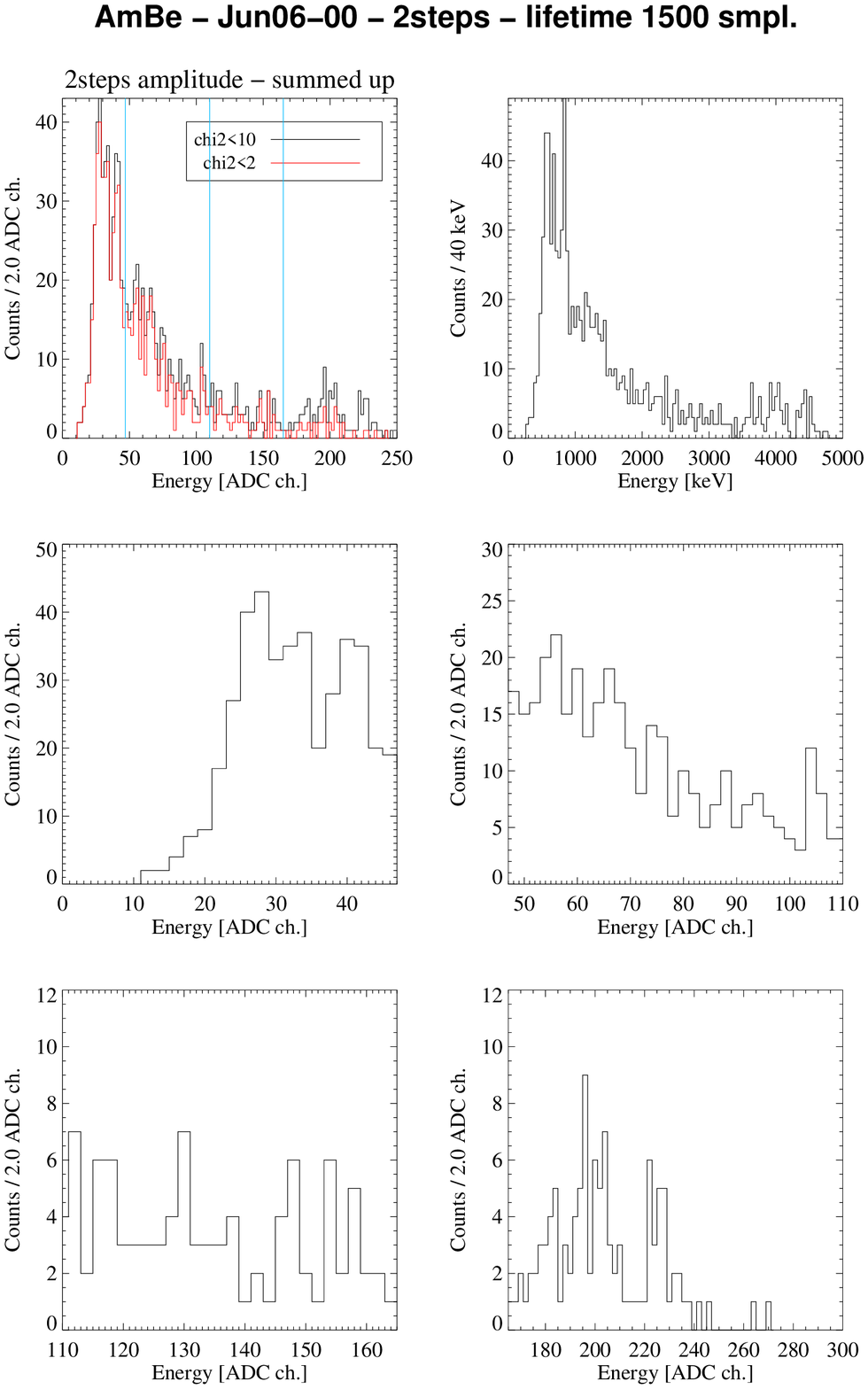}
\includegraphics[bb=310 526 522 732,clip]{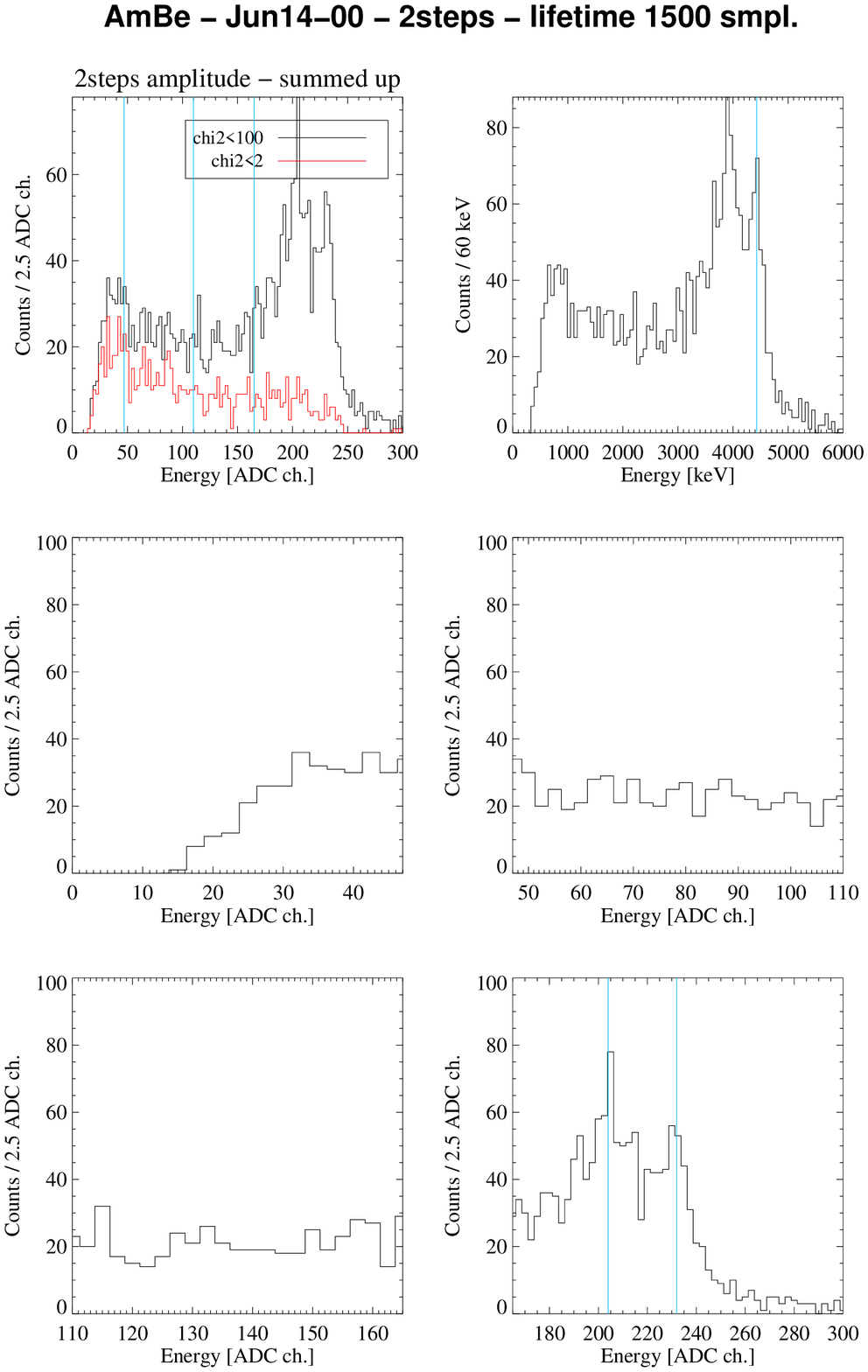}
\caption{\label{f:AmBe:spec}
 \emph{Left:} Am-Be spectrum taken with light trigger settings during the flight.
 \emph{Right:} Am-Be spectrum taken with settings optimized for MeV energies.}
\end{figure}

\begin{figure}
\centering
\includegraphics[bb=70 68 518 740,width=0.7\textwidth,clip]{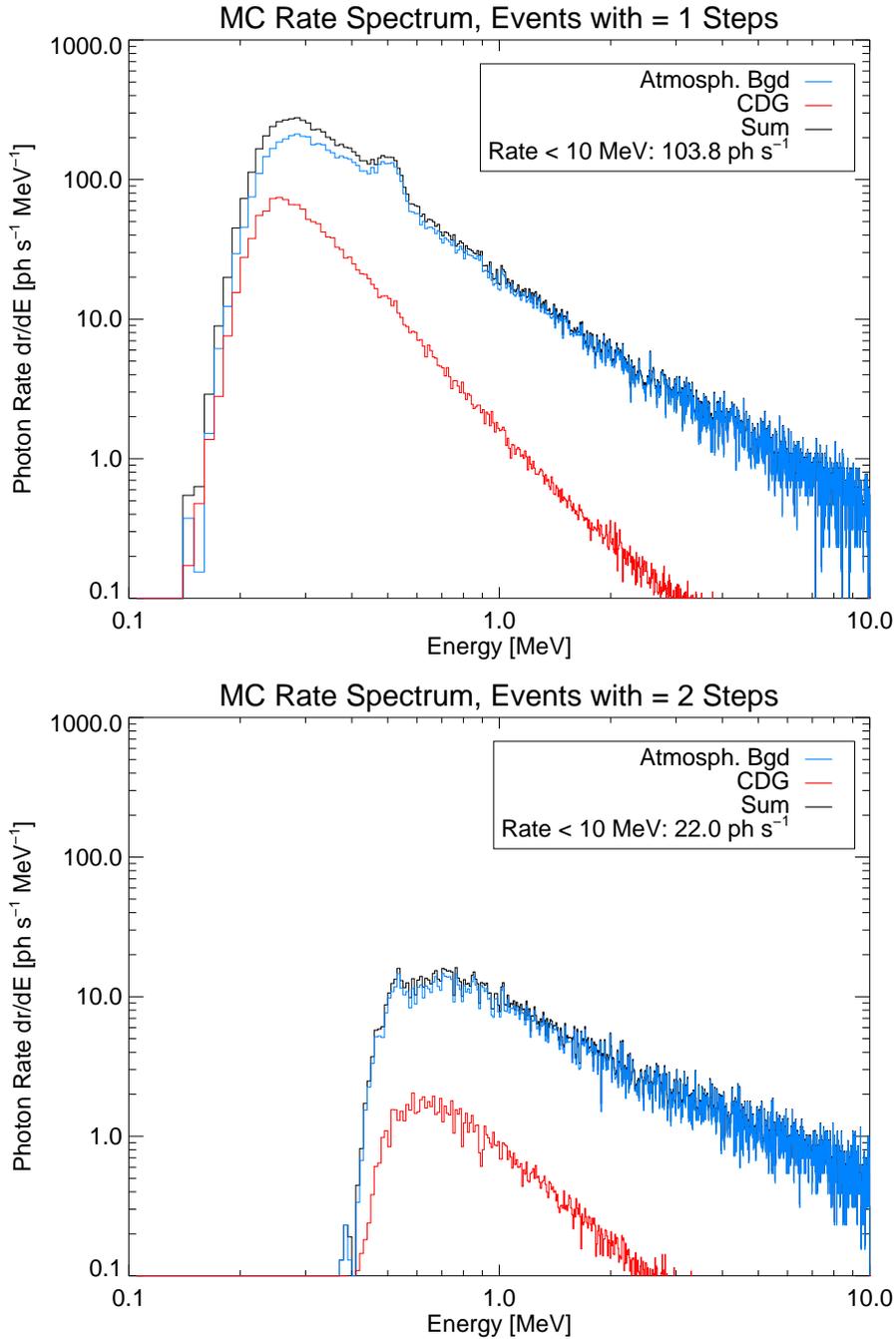}
\caption{\label{f:mc:atmcdg}
 Expected spectrum from Monte Carlo simulations for \g-rays from atmospheric
 background and cosmic diffuse \g-ray background (CDG) in the energy range from
 0.1 -- 15~\MeV, interacting within the sensitive LXe volume. The adapted
 instrumental response is simplified. 
}
\end{figure}
\subsection{Preliminary Results from Monte Carlo Data}
\label{s:simul}
To fully understand the measured background rate in LXeGRIT in the
May 1999 flight, we need to finalize the analysis of all the data and we need to study further all possible contributions to the measurement. Monte Carlo simulations of the detector's response to a variety of background sources are under way, using a detailed mass model of the instrument within the simulation package GEANT \cite{GEANT:00}~. The estimated contribution to the \g-ray background from the atmosphere and from the
cosmic diffuse background is  about 160~Hz in the sensitive LXe volume, increasing to 650~Hz
if interactions in the passive LXe are included. These estimates are for events with a minimum energy deposition of 100~\keV\ in each interaction. 
For the contribution from the cosmic diffuse
\g-ray background, mainly entering through the instrument's aperture, we used
the flux given by \cite{VSchoenfelder:80}~. For the dominant atmospheric
background, we used an approximated continuum description of the flux by
\cite{ECosta:84:atm_bgd}~, extrapolating the given fluxes down to 100~\keV\
and up to 15~\MeV. The resulting spectra of \g-ray interactions are shown in
Fig.~\ref{f:mc:atmcdg} for single-step (top panel) and multiple-step events
(bottom panel). These Monte Carlo simulations have taken into account the measured LXeTPC energy and position resolutions, 
as well as a minimum  energy threshold, which for the results shown in Fig.~\ref{f:mc:atmcdg}, was taken as 150~keV. They did not take into account, however, trigger efficiency and its energy dependence, data taking and reconstruction
efficiencies. The impact of these selections has to be studied before a meaningful comparison with the flight data, since  both the spectral shape and the contribution of events with single and multiple interactions are affected.
It is clear, however, that the observed ratio of single to multiple interaction events is much higher than predicted from the simulations of atmospheric and diffuse background.

\begin{figure}[p]
\centering
\includegraphics[width=\textwidth,clip]{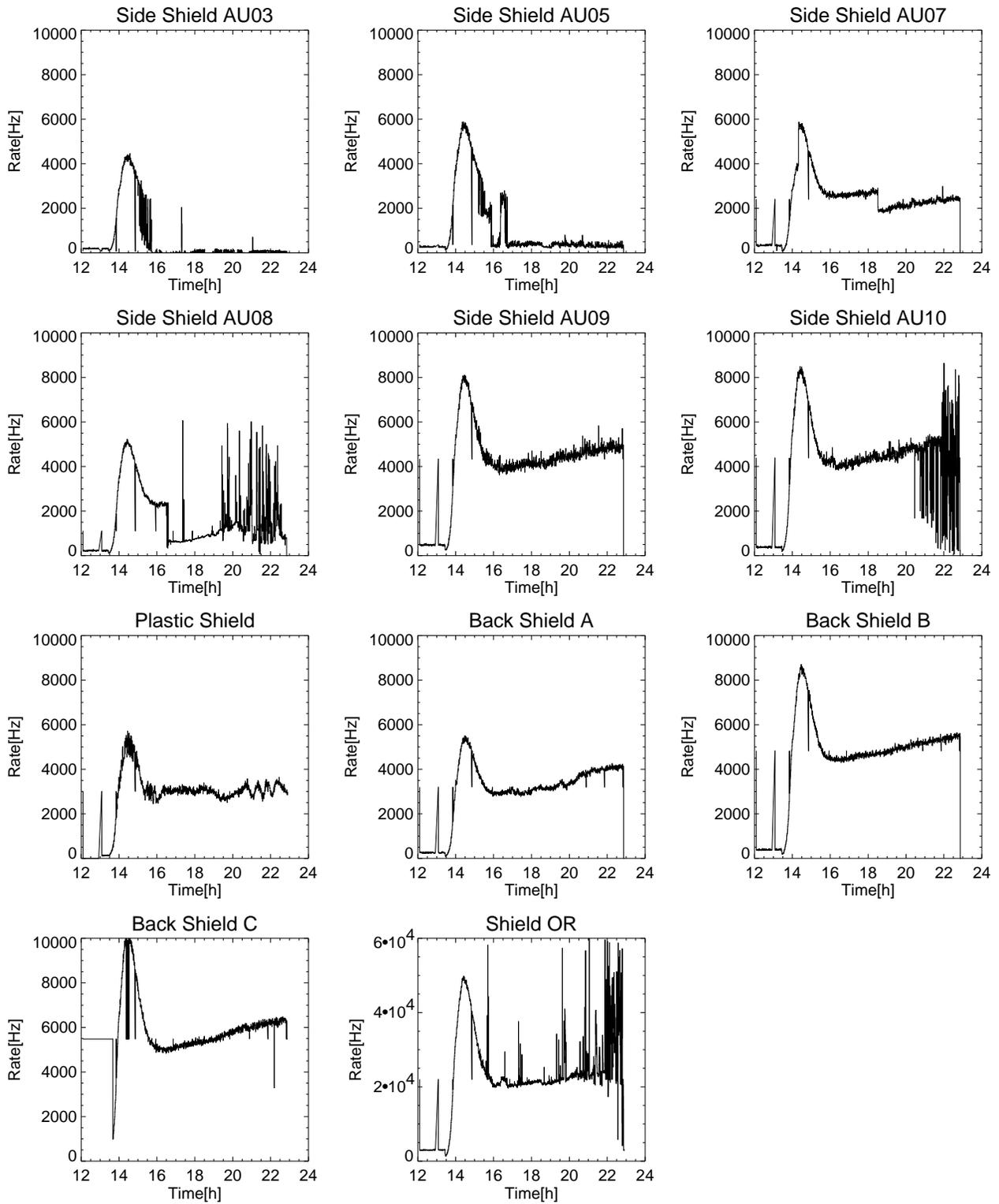}
\caption{\label{f:vetorates}
 Veto rates from the six side shield units and the three back
 units. The lower two plots refer to the  plastic shield rate and the logical  OR of all the units.
}
\end{figure}

It is well known that \g-rays other than from atmospheric and diffuse background cannot be neglected. These notably include background \g-rays and beta decays from activity in all instrument materials initiated by albedo protons and neutrons as well as by locally produced neutrons.
To study the impact of hadron interactions in the LXeGRIT payload, we have modified our Monte Carlo simulations code to include the transport of neutrons.  The mass model, previously developed for the \g-ray
simulations, has been refined to reflect the presence of all materials relevant for neutron interactions.
 We use
the GEANT package together with GCALOR, an interface to the CALOR hadron
simulation package \cite{GCALOR:00:neu}~.  Atmospheric neutrons from thermal
energies up to 100~\MeV\ have been propagated through the payload, registering
all neutron capture reactions. Measured atmospheric neutron fluxes and
distributions have been collected from the literature, e.g. \cite{Preszler:76:neutrons}~.  As described in Section~\ref{s:veto}, the back
shields consist of pure NaI crystals, while the side shields consist of crushed
NaI immersed in a liquid scintillator, a good neutron moderator. Since two xenon isotopes (out of nine stable isotopes) have large thermal neutron
cross sections, namely $\sim 100$~b for $^{131}$Xe with an abundance of 21.2\%
and $\sim 22$~b for $^{129}$Xe (26.9\%), production of thermal neutrons around
the LXeTPC is of concern. It turns out, however, that the fraction of neutron
captures within the xenon (34~Hz in the active volume and 89~Hz in the passive
volume) is relatively small compared to the total estimated capture rate of 3.4~kHz. This rate
is dominated by captures within the shield system, with $\sim2.4$~kHz in the side shields and  $\sim 550$~Hz in the back shields. The remaining
capture rate is found in the stainless steel body of the detector. Work is
currently underway to model the \g-ray cascades and radioactive decays following
the captures, in order to ultimately assess the rate and spectrum of \g-rays
expected in the chamber. We have also started to study fast neutron reactions
such as elastic neutron scatters, (n,p)-reactions, etc.

In fact, the rate on all the shield units, shown in Fig.~\ref{f:vetorates}, gradually increase
with time. This effect is apparently consistent with activation of NaI, and with
the build-up of radioactive nuclei such as $^{24}$Na with a halflife of 15~h.
In addition, three sections of the side shields showed excessive rates at float
altitude, which required to raise their thresholds. This rendered a large part
of the side shield essentially passive below several 100~\keV, such that
a significant number of low-energy gamma-rays may have leaked into the LXeTPC
without veto. Also shown in the same Fig.~\ref{f:vetorates}, is the rate measured in the plastic scintillation counter above the LXeTPC. The average rate which remains constant to about 3kHz, at float altitude, is consistent with the expected cosmic rays rate on this 1600 cm$^2$ counter. This number also agrees with the estimate obtained from subtracting the TPC rate within the discriminator window from the total rate above the lower discriminator threshold.

\section{Conclusions}

The May 7, 1999 balloon flight of LXeGRIT verified a good performance of this instrument in the near space environment. A total of about 300000 events were accumulated during the 9~h flight. The
presumably fragile LXeTPC structure and its support systems not only survived the
flight, but also the rough landing conditions with a parachute and the 350 miles 
transport back to the launch base. The detector, still filled with LXe, showed a similar signal response on all wires and anodes as before launch.  Analysis of the data from the May '99 flight, while still in
progress, shows that the main goal of the background measurement at balloon altitude in this new Compton telescope has been achieved. The in-flight background spectrum is dominated by single interaction events below about 600 keV, partly due to the trigger conditions used for the flight data taking. The impact of all possible sources to the measured background is being studied with Monte Carlo simulations using a realistic instrument mass model.   

\acknowledgments  

We would like to thank the NSBF team for excellent support throughout the flight campaign. Support for the LXeGRIT project is provided by NASA High Energy Astrophysics Division SR\&T program under grant NAG5-5108.

\end{document}